\documentclass[twocolumn]{aastex7}

\usepackage[utf8]{inputenc}
\usepackage{hyperref}
\usepackage{acronym}   
\usepackage{xspace}
\usepackage{amsmath}
\usepackage{amssymb}
\usepackage{xcolor}
\usepackage{graphicx}
\hypersetup{pdfauthor={Name}}

\newcommand{\Msun}{$\mathrm{M}_\odot$}

\acrodef{DNS}{double neutron star}
\acrodef{GRB}{gamma-ray burst}
\acrodef{SN}{supernova}

\shortauthors{Mandel et al.}
\shorttitle{Merger GRB offsets}

\begin{document}

\title{The maximum offsets of binary neutron star mergers from host galaxies}
\date{\today}

\correspondingauthor{Ilya Mandel}

\author[0000-0002-6134-8946]{Ilya Mandel}
\affiliation{School of Physics and Astronomy
Monash University,
Clayton, VIC 3800, Australia}
\affiliation{The ARC Centre of Excellence for Gravitational Wave Discovery -- OzGrav, Australia}
\email{ilya.mandel@monash.edu}

\author{Om Sharan Salafia}
\affiliation{INAF -- Osservatorio Astronomico di Brera, via Brera 28, I-20121 Milano (MI), Italy}
\affiliation{INFN -- sezione di Milano-Bicocca, Piazza della Scienza 3, I-20126 Milano (MI), Italy}
\email{om.salafia@inaf.it}

\author{Andrew Levan}
\affiliation{Department of Astrophysics/IMAPP, Radboud University, 6525 AJ Nijmegen, The Netherlands}
\affiliation{Department of Physics, University of Warwick, Coventry, UK}
\email{a.levan@astro.ru.nl}

\author[0000-0002-0492-4089]{Paul Disberg}
\affiliation{School of Physics and Astronomy
Monash University,
Clayton, VIC 3800, Australia}
\affiliation{The ARC Centre of Excellence for Gravitational Wave Discovery -- OzGrav, Australia}
\email{paul.disberg@monash.edu}

\begin{abstract}
We analytically derive, and illustrate with a population synthesis model, the maximum offset of binary neutron star mergers ejected from their host galaxies.  This approximate maximum offset is 300 kpc $\times\ (v_\mathrm{esc} / 500\ \mathrm{km}\ \mathrm{s}^{-1})^{-7}$, where $v_\mathrm{esc}$ is the escape velocity from the host galaxy.  Massive hosts with high escape velocities are unlikely to yield very large offsets.  This maximum offset should inform the host associations of mergers that are not coincident with galaxies.  We also discuss potential correlations between offsets and system masses, and possibly the duration of the gamma-ray burst accompanying the merger.
\end{abstract}


\section{Introduction}
\label{sec:intro}

Some 30\% of merger-origin \acp{GRB} appear not to lie directly on top of any galaxy, with the merging binary presumably escaping from the host galaxy with a large systemic velocity $v_\mathrm{sys}$  \citep{BloomProchaska:2006, Berger:2010,Tunnicliffe:2013}.  The association of these hostless \acp{GRB} to potential hosts has significant implications for the exploration of binary evolution and supernova physics, including in the context of gravitational-wave astronomy \citep[e.g.,][]{WandermanPiran:2015,Kelley:2010}.  At present, the standard methodology for this association relies on estimating the probability of a chance projected alignment between the \ac{GRB} location and the host galaxy, which favours larger offsets from rare massive hosts \citep{Bloom:2002,Levan:2007,Berger:2010,Gaspari:2025}.   

Merger-origin \acp{GRB} are likely dominated by merging \acp{DNS}, since black holes would need to have a low mass, a rapid spin, or both, in order to disrupt the neutron star companion during a merger of a neutron star with a black hole \citep{Foucart:2018}.  We will therefore focus on \ac{DNS} mergers.

Achieving a significant offset from the host galaxy requires both a large systemic velocity  of the binary, in order to escape from the galactic gravitational potential well, and a sufficient delay time prior to merger, set by the timescale $\tau_\mathrm{GW}$ of coalescence due to gravitational-wave emission \citep{Peters:1964}.

As we will argue below, for the majority of \acp{DNS}, both the systemic kick and the coalescence time are determined by orbital properties before the second \ac{SN}.  The systemic velocity is capped by the orbital velocity before the second supernova $v_\mathrm{orbPreSN2}$, so tighter binaries can achieve larger systemic velocities.  However, the coalescence time is a much stronger function of the separation, so very tight binaries merge too quickly to achieve significant offsets.  The largest offsets are thus determined by the escape speed $v_\mathrm{esc}$ from the host galaxy.  

We carry out these simple calculations and illustrate them with results from population synthesis models in section 2.  In section 3, we discuss some of the implications and caveats, including the apparent implausibility of some of the large offsets claimed from massive hosts.  We also comment on the possibility that the highest systemic kicks may correspond to a particular range of compact object masses, which may in turn determine the electromagnetic signature from the merger, potentially leading to a correlation between the \ac{GRB} type, the host offset, and the delay time.

\section{Maximum offsets}

We begin with relatively confident, though approximate, statements, which we subsequently illustrate with population synthesis results obtained with the COMPAS rapid binary population synthesis code \citep{Stevenson:2017,VignaGomez:2018,COMPAS:2021,COMPAS:2025}.

\textbf{1. The maximum systemic kick of a \ac{DNS} is set by the orbital velocity of the binary before the second supernova.}

The survival and post-supernova orbital properties of a circular binary are derived by, e.g., \citet{BrandtPodsiadlowski:1994, TaurisTakens:1998}.  In general, these depend on the fraction of mass lost in the supernova and the magnitude and direction of a natal kick received by the newly formed compact remnant.  These quantities are uncertain and may be correlated \citep[e.g.,][]{Fryer:2012,MandelMueller:2020,Valli:2025}.  We therefore opt for general arguments rather than attempting to compute a distribution of systemic kicks over a particular distribution of progenitor properties, mass loss fractions, and kick magnitudes and directions.

An asymmetric natal kick significantly larger than the pre-supernova orbital velocity is overwhelmingly likely to disrupt the binary.  In the absence of mass loss, the largest kick that can keep a circular binary intact has a magnitude of $(\sqrt{2}+1)$ times the pre-supernova orbital velocity and is directed opposite to the initial orbital motion; such a natal kick would give an equal-mass binary a systemic kick that is $(\sqrt{2}+1)/2 \approx 1.21$ times the pre-supernova relative orbital velocity.  Maximum systemic kicks are an even lower multiple of the pre-supernova orbital velocity at periapsis for initially eccentric binaries.   Meanwhile, the \citet{Blaauw:1961} kick associated with symmetric mass loss is limited to a fraction of the pre-supernova orbital velocity; assuming that the binary will have equal masses after the supernova (a \ac{DNS}, see below), the \citet{Blaauw:1961} kick velocity $v_\mathrm{Blaauw2}$ is strictly limited to no more than 50\% of the pre-supernova orbital velocity for surviving binaries in the absence of natal kicks.  While the exact value of the systemic kick depends on the combination of the natal kick magnitude and direction and the amount of mass lost, the total systemic kick imparted on the \ac{DNS} by the second supernova cannot exceed the pre-supernova orbital velocity by more than 25\% for binaries that are initially circular and that have equal final masses. 
 
In the standard channel of \ac{DNS} formation \citep[e.g.,][]{Tauris:2017,VignaGomez:2018,Stevance:2023} a binary with $\approx 8$--$15$ \Msun\, components starts out rather wide, with an initial separation of several hundred solar radii.  The primary overflows its Roche lobe after it evolves off the main sequence, as a Hertzsprung gap or core-helium burning star; the first episode of mass transfer is mostly non-conservative and keeps the binary wide.  The first supernova thus happens in a wide binary, with a typical orbital velocity of $\lesssim 75$ km s$^{-1}$.   Thus, if the wide binary is to survive the first supernova, the primary's natal kick must be low.  For example, the primary may explode in an electron-capture supernova \citep{BeniaminiPiran:2016,Willcox:2021} and the systemic kick due to this first supernova rarely exceeds 100 km s$^{-1}$.  This is unlikely to eject the binary even from low-mass galactic potentials.

Mass transfer from the evolving secondary to the neutron-star primary can be markedly different because, following the first supernova, the primary's mass ($\sim 1.3$ \Msun\ for typical Galactic \ac{DNS}) is much lower than that of the secondary ($\gtrsim 8$ \Msun).  This large mass ratio means that the reverse mass transfer is very likely to be dynamically unstable.  The resulting common-envelope phase \citep[see][for reviews]{Ivanova:2013,RoepkeDeMarco:2023} can harden the binary by several orders of magnitude, to a separation of only a few solar radii (it cannot be too small since the secondary's core, a naked helium star, must still fit into the binary).  It is frequently assumed that the binary will become approximately circular following the common-envelope phase, though torquing by a circumbinary disk, if one persists, could drive up eccentricity \citep{Wei:2023}.  The orbital velocity at the time of the second supernova can reach some 600 km s$^{-1}$.  Therefore, by the same argument as above, systemic kicks reaching 800 km s$^{-1}$ are possible from a combination of a natal kick and Blaauw kick accompanying the second supernova.\footnote{Although observed Galactic DNSs are estimated to have received systemic kicks of $\lesssim$ 100 km s$^{-1}$ \citep{Disberg:2024}, the existence of higher systemic kicks cannot be excluded because selection effects would likely render them unobservable \citep{Disberg:2025}.}

\begin{figure}[!tb]
\centering
\includegraphics[width=0.5\textwidth]{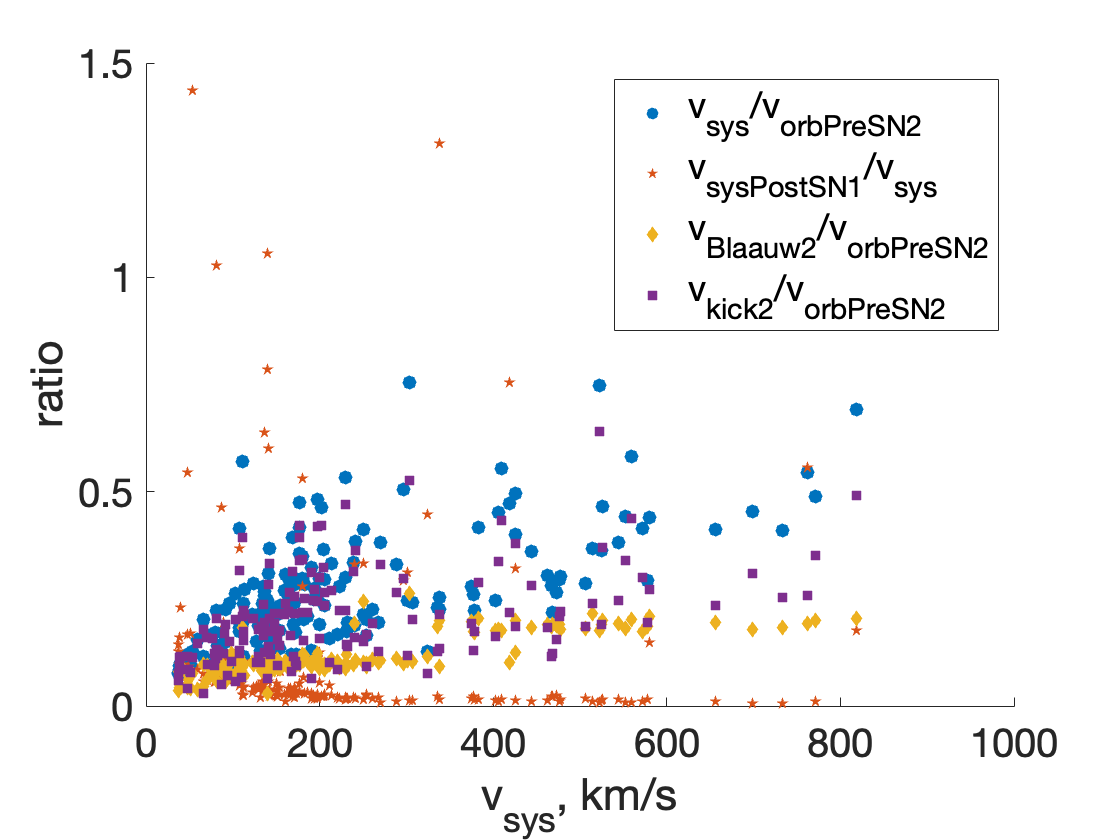}
\caption{The final systemic kick velocity $v_\mathrm{sys}$ of merging \acp{DNS} is limited by the orbital velocity prior to the second supernova (blue circles), with only a small subset of binaries receiving significant contributions from the first supernova (red stars); contributions from \citet{Blaauw:1961} (yellow diamonds) and natal (purple squares) kicks associated with the second supernova are also shown for this COMPAS simulation.} 
\label{fig:vel}
\end{figure}

We illustrate these points in Figure \ref{fig:vel} with the results of a COMPAS rapid binary population synthesis simulation.  The simulation used default settings for version v3.22.06 \citep[see][]{COMPAS:2025}, including supernova remnant masses and kicks taken from the \citet{MandelMueller:2020} prescription.  The simulation was carried out at solar metallicity.  Only \acp{DNS} merging within 14 Gyr are shown.  The larger blue dots indicate the ratio of the final systemic kick to the relative orbital velocity just before the second supernova.  Although this value can in principle reach $\approx 1.25$, such extreme systemic kicks are rare, and the ratio does not exceed 1 for any of the 166 simulated merging \acp{DNS} plotted in Figure \ref{fig:vel}.  The yellow and purple dots break this down into contributions from the \citet{Blaauw:1961} and asymmetric natal kicks associated with the second supernova, respectively.  The red dots indicate that the contribution of the first supernova to the final systemic kick is generally small, with the exception of a few cases (mostly at lower systemic kicks, and typically connected to binaries that already have a short period prior to the first supernova, such as those experiencing double-core common-envelope events), justifying our focus on the second supernova.

\textbf{2. The maximum delay time between this kick and the \ac{DNS} merger is also set by the orbital velocity of the binary before the second supernova.}

The delay time $\tau_\mathrm{GW}$ due to gravitational-wave emission for a circular binary with separation $a$ and equal component masses $M$ is given by \citet{Peters:1964}:
\begin{equation}
\tau_\mathrm{GW} =  550\, \mathrm{Myr} \left(\frac{M}{1.3\, \mathrm{M}_\odot}\right)^{-3}  \left(\frac{a}{2\, \mathrm{R}_\odot}\right)^{4}\, .
\end{equation}
Meanwhile, the orbital velocity scales as $\sqrt{GM/a}$.  Thus, $\tau_\mathrm{GW} \propto a^4 \propto v^{-8}$.  The maximum distance should then be bounded by $v \times \tau_\mathrm{GW} \propto v^{-7}$, where $v$ is the orbital velocity before the second supernova, since the systemic kick is at most slightly greater than $v$; furthermore, the time to merger cannot be longer than the age of the Universe, placing an absolute bound of $v \times 14$ Gyr on the maximum distance.

Of course, the second supernova explosion can make the binary eccentric.  The gravitational-wave merger timescale is longer for an eccentric binary than for a circular binary with the same periapsis distance $r_\mathrm{p}$, scaling as $\tau_\mathrm{GW} \propto r_\mathrm{p}^4 (1-e)^{-1/2}$ \citep{Peters:1964} in the limit $e \to 1$.  A supernova can increase the semi-major axis of a circular binary; however, the periapsis stays the same or shrinks.  In practice, $\tau_\mathrm{GW} \propto v^{-8}$ remains a good approximation for the maximum delay time, provided the systemic kick $v_\mathrm{sys}$ rather than the orbital velocity is used as $v$, except in very rare cases with fine-tuned properties, such as natal kicks whose magnitude and direction yield near-unity eccentricities to surviving binaries \footnote{Under specific kick model assumptions, the merger timescale distributions and hence the offset distributions can be computed explicitly \citep[e.g.,][]{BeniaminiPiran:2024,VignaGomez:2025}.}.  We thus approximate the maximum distance $D_\mathrm{max}$ from the birth site which the \ac{DNS} binary can reach before merging as
\begin{eqnarray}
\label{eq:Dmax}
D_\mathrm{max} &=& v_\mathrm{sys} \times \min\left(14\ \mathrm{Gyr}, \tau_\mathrm{GW}\right)\\
&\approx& 5\ \mathrm{Mpc} \times \min\Bigg(\frac{v_\mathrm{sys}}{335\ \mathrm{km}\ \mathrm{s}^{-1}},
	\left( \frac{v_\mathrm{sys}}{335\ \mathrm{km}\ \mathrm{s}^{-1}}\right)^{-7}\Bigg). \nonumber
\end{eqnarray}
Here, we assumed fiducial component masses of $\sim 1.3\, \mathrm{M}_\odot$, consistent with the peak of the observed distribution of Galactic \acp{DNS} \citep{Farrow:2019}.

\begin{figure}[!tb]
\centering
\includegraphics[width=0.5\textwidth]{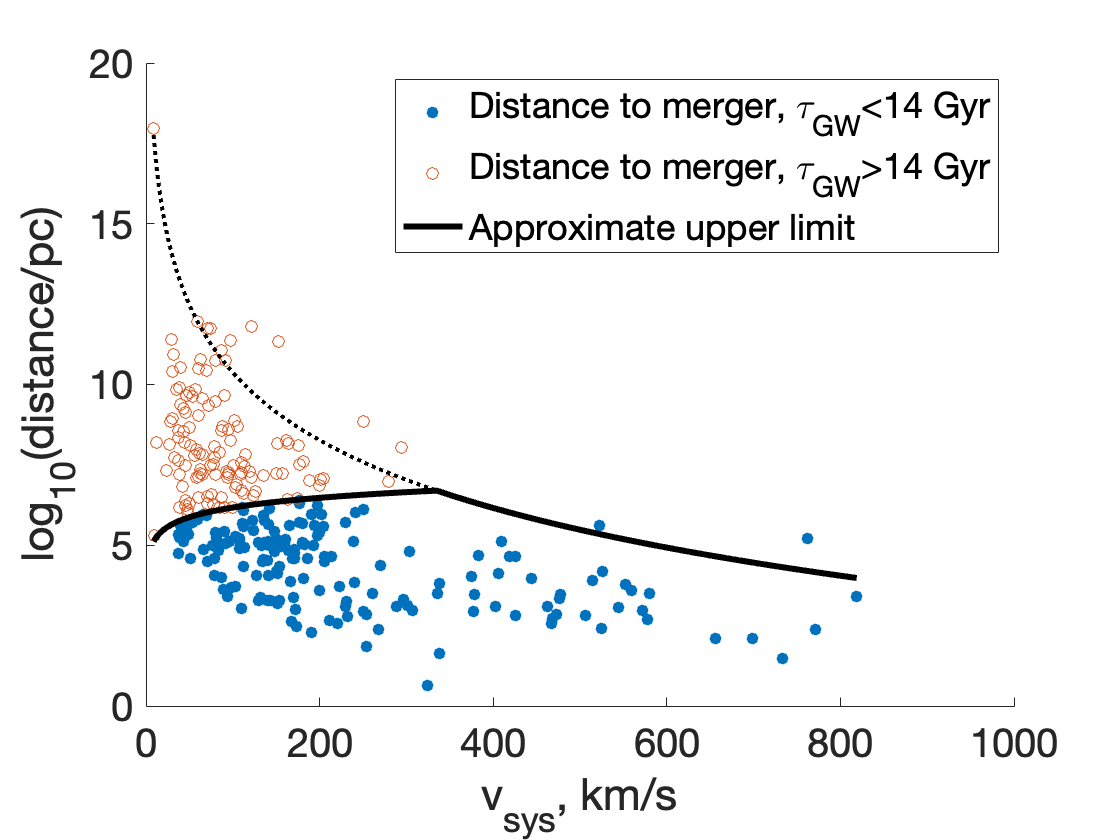}
\caption{The distance travelled by merging \acp{DNS} from the COMPAS simulation as a function of the binary systemic kick, ignoring the impact of the galactic potential.  Filled blue and open red circles indicate \acp{DNS} that merge within 14 Gyr or later, respectively.  The solid black line indicates the maximum distance estimate of Equation \ref{eq:Dmax}, while the dotted black line indicates its second term only, without the constraint that mergers must happen within 14 Gyr.  Approximately 99\% of merging \acp{DNS} do so 
within the maximum distance estimate of Equation \ref{eq:Dmax}.} 
\label{fig:delay}
\end{figure}

Figure \ref{fig:delay} shows the distances travelled between \ac{DNS} formation and merger by binaries in our COMPAS simulation, in the absence of any post-formation gravitational interactions.  While it is possible for some binaries to have longer delay times and hence longer travel distances than estimated in Equation \ref{eq:Dmax}, $\approx99$\% of merging binaries satisfy this maximum distance estimate.
 
\textbf{3. The maximum offset for a merging \ac{DNS} ejected from a host galaxy is set by the host galaxy's escape velocity.}

We have shown in point {\bf 1} that the maximum distance travelled by the binary scales as $v^{-7}$ when ignoring the constraint that binaries should merge within the current age of the Universe.  Equation \ref{eq:Dmax} and Figure \ref{fig:delay} show that, with the inclusion of this constraint, the largest offsets could be obtained by binaries with relatively low systemic kicks of 335 km s$^{-1}$.

However, our discussion so far has ignored the gravitational effects of the host galaxies.  Binaries with systemic kicks appreciably lower than the host's escape velocity $v_\mathrm{esc}$ (i.e., smaller orbital velocities than $v_\mathrm{esc}$ before the second supernova, according to point {\bf 1}) will not be able to travel significantly beyond the host galaxy's halo.  The systemic kicks yielding the highest offsets are therefore those with kicks slightly exceeding $v_\mathrm{esc}$.    

Thus, the maximum distance travelled is  
\begin{equation}
D_\mathrm{max} \lesssim \begin{cases}
	300 \, \mathrm{kpc}\left(\frac{v_\mathrm{esc}}{500\, \mathrm{km}\, \mathrm{s}^{-1}}\right)^{-7} & \text{if}\ v_\mathrm{esc} > {335\, \mathrm{km}\, \mathrm{s}^{-1}}\\
	5\, \mathrm{Mpc}  & \text{otherwise.}
\end{cases}
\label{eq:Dmaxesc}
\end{equation}
This estimate does not account for the initial velocity of the binary within the host galaxy.  However, since massive stars are generally born deep inside the galactic halo, the local velocity is typically small with respect to the total systemic velocity of the binary; for example, the circular rotational velocity at the Sun's location is some 230 km s$^{-1}$ while the escape velocity from the same location is around 550 km s$^{-1}$ \citep{KoppelmanHelmi:2021}.  Moreover, the upper limit in Equation \ref{eq:Dmaxesc} is already conservative, since the binary will slow as it climbs out of the host galaxy's potential, reducing the offset by an amount that depends on the details of the birth location and the host galaxy's potential.  On the other hand, if the host galaxy experienced one or more major galactic mergers between the time of the \ac{DNS} ejection and \ac{DNS} merger, the galactic potential may have evolved significantly and the escape velocity at the time the \ac{DNS} was originally ejected could have been significantly lower than the current value \citep[e.g.,][]{Kelley:2010,PeretsBeniamini:2021}.

\section{Discussion}

The standard methodology for determining the likely host association of apparently hostless \ac{GRB} afterglows favours rare luminous massive galaxies at larger offsets over low-luminosity common galaxies closer to the source's location \citep{Bloom:2002,Levan:2007,Berger:2010,Gaspari:2025}.  

However, massive galaxies are expected to have higher escape velocities.  The relationship between galactic masses and radial extents suggests that the rotational velocities scale as $v \propto M^{1/4}$ \citep{TullyFisher:1977,Schulz:2017}.  If we assume that the typical escape velocity scales with the rotational velocity, and the mass scales with the galactic stellar mass $M_*$ (i.e., ignoring complex relationships between stellar mass, baryonic mass, and dark matter halo mass), we can rescale the escape velocity to the Milky Way solar  neighborhood $v_\mathrm{esc} = 550$ km s$^{-1}$.  For a Milky Way stellar mass of $3\times 10^{10}$ M$_\odot$ \citep{Lian:2025}, this yields $v_\mathrm{esc} \approx 750\ (M_*/10^{11} \mathrm{M}_\odot)^{1/4}$ km s$^{-1}$, so Equation \ref{eq:Dmaxesc} becomes
\begin{equation}
D_\mathrm{max}  \lesssim \begin{cases}
	17\, \mathrm{kpc} \left(\frac{M_*}{10^{11} \mathrm{M}_\odot}\right)^{-7/4} &  \text{if}\ M_* > 4\times10^9  \mathrm{M}_\odot\\
	5\, \mathrm{Mpc}  & \text{otherwise.}
\end{cases}
\label{eq:Dmaxmass}
\end{equation}

\begin{figure}[!tb]
\centering
\includegraphics[width=0.5\textwidth]{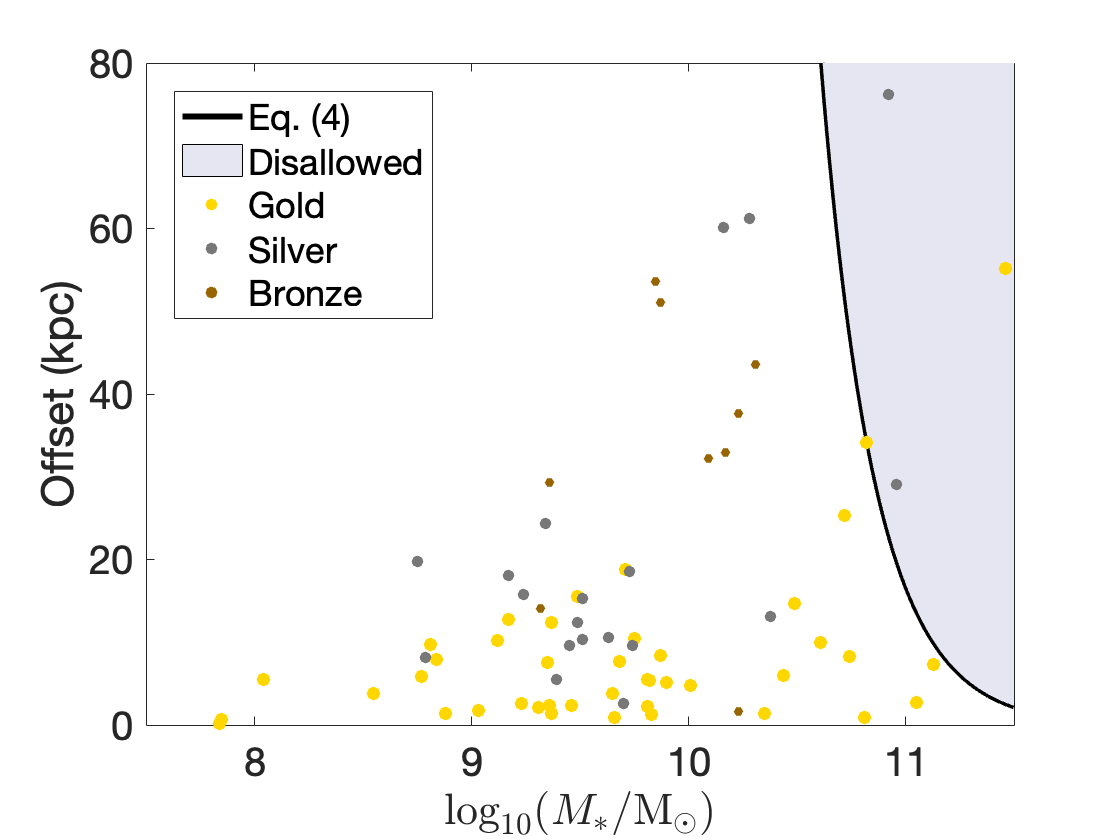}
\caption{The claimed host offsets of short \acp{GRB} from the samples of \citet{Fong:2022,Nugent:2022} as a function of putative host stellar mass, divided into Gold (large gold dots), Silver (medium silver dots), or Bronze (small bronze dots) based on the chance alignment probability following current host association protocols.  Equation \ref{eq:Dmaxmass} is shown in black, with the region of disallowed offsets above it shaded.} 
\label{fig:observed}
\end{figure}

Figure \ref{fig:observed} shows the offsets of claimed host galaxy associations against putative host stellar masses from the short \ac{GRB} sample of \citet{Fong:2022,Nugent:2022}.  This does not include mergers associated with long \acp{GRB}.  The associated offsets are split up into Gold, Silver, and Bronze according to the current estimates of chance alignment probability.  The curve of Equation \ref{eq:Dmaxmass} is shown in black, with the disallowed region above it shaded in grey.  

The constraint developed here is most useful for high host galaxy masses.  There, large kicks are required to escape the potential, so any ejected binary must be very tight and would therefore merge promptly through gravitational-wave emission, before it could reach large offsets.  At least one Gold association, GRB 050509B, is clearly ruled out (although we note this is in a galaxy cluster environment, where dynamics may be more complex -- \citealt{Gehrels:2005}), while another, GRB 070809, appears unlikely.  

In practice, the maximum offset should not be taken as a hard boundary, but as a probabilistic statement, after accounting for the distribution of escape velocities depending on the kick location, the deceleration of escaping binaries moving through the galaxy's potential, and the angular dependence of the projected offset, among other factors.  We expect that some of these considerations may make the host associations unlikely even for some of the events that appear in the white (allowed) region in Figure   \ref{fig:observed}.  This motivates a revision to the procedure of host galaxy association.  

While this conclusion is robust to assumptions about the details of natal kicks, we now turn to more tentative considerations.  

Despite very significant progress in understanding the formation of neutron stars in supernovae, including the asymmetric kicks they receive at birth, significant challenges remain (see \citealt{Popov:2025} for a recent review).  One interesting possibility for surpassing the limit derived here could be continuous acceleration by so-called ``rocket'' kicks; although these are unlikely to significantly affect systemic velocities, they may increase the periapsis separations of strongly kicked, and therefore highly eccentric, binaries, increasing their merger times and offsets \citep{Hirai:2024}.

\begin{figure}[!tb]
\centering
\includegraphics[width=0.5\textwidth]{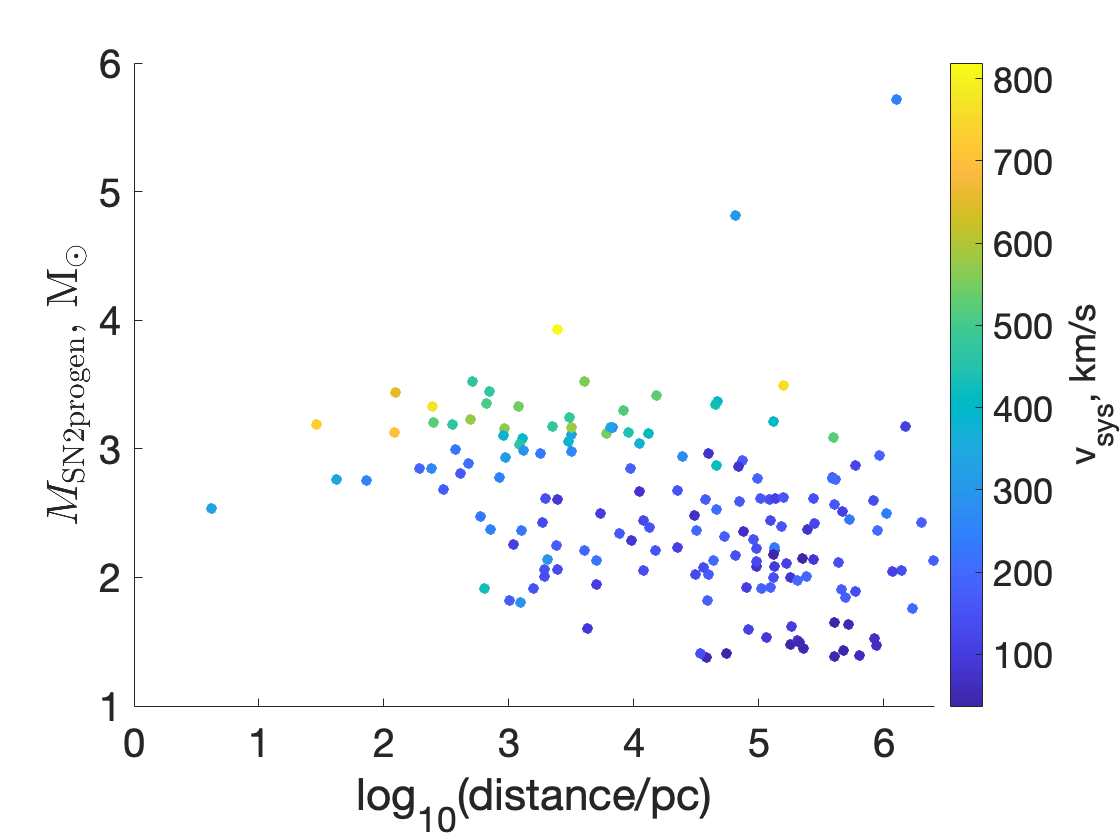}
\caption{The pre-supernova mass of the progenitor of the second-born neutron star vs.~the distance travelled prior to merger (ignoring the impact of the galactic potential), for \acp{DNS} merging within 14 Gyr.  The color denotes the systemic kick of the binary.} 
\label{fig:offset}
\end{figure}

For illustration purposes, we chose one particular model of neutron star remnant masses and natal kicks \citep{MandelMueller:2020}.  While this model is able to reproduce some observations, such as the velocity distribution of single pulsars (\citeauthor{Kapil:2022} \citeyear{Kapil:2022}, cf.\ \citeauthor{DisbergMandel:2025} \citeyear{DisbergMandel:2025}), it is in tension with other data, including the period-eccentricity distribution of Galactic \acp{DNS} \citep{Mandel:2020}.  This model (stochastically) predicts both the remnant mass and the remnant kick on the basis of the properties of the progenitor.  While the exact details of the prescription may well prove incorrect, the natural coupling between masses and kicks is appealing (see also models of \citealt{BrayEldridge:2016,BrayEldridge:2018}).  

Figure \ref{fig:offset} demonstrates one consequence of this mass--kick coupling: most of the binaries which receive the largest systemic kicks are ones with high progenitor masses, above 3 M$_\odot$.  These are the systems likely to escape from the host galaxy's potential: although the systems in bottom right of Figure \ref{fig:offset} have larger nominal distances travelled, their systemic velocities are too small to escape from all but the lightest hosts.  Almost all \ac{DNS} binaries that merge within 14 Gyr are sufficiently compact that they experience ultra-stripping prior down to the carbon-oxygen core prior to the core collapse of the secondary \citep{Tauris:2015} in the COMPAS models.  Within the \citet{MandelMueller:2020} model, these $>3$ M$_\odot$ naked carbon-oxygen cores create a specific, relatively narrow range of neutron star masses around $1.43 \pm 0.06$ M$_\odot$.  Meanwhile, the primaries in these binaries typically experience electron-capture supernovae \citep{Nomoto:1984}, leaving behind $\approx 1.26$ M$_\odot$ neutron stars, although these may grow by $\sim 0.1$ M$_\odot$ through subsequent accretion.  

Taken at face value, this model suggests that significantly offset, hostless binary mergers may have a specific range of \ac{DNS} masses, slightly above the average for known merging Galactic \acp{DNS}.  While the overall differences in mass are small, the total masses of most merging binary neutron stars may scatter in the vicinity of 1.2 times the Tolman-Oppenheimer-Volkoff limit (depending on the equation of state), so even small changes could impact whether the merger product is a supermassive neutron star, a hypermassive one, or promptly collapses into a black hole \citep{Salafia:2022}.  This in turn could affect the electromagnetic signature of the merger.  The total mass and mass ratio may also impact the survival duration of the accretion disk around the merger remnant and the associated electromagnetic emission  \citep{Gottlieb:2023}.  Whatever the exact mechanism, it is natural to assume that the component masses of a merging \ac{DNS} determine, for example, the duration of the \ac{GRB}, if any, accompanying the merger.

Thus, the relatively large offsets of some merger-associated long GRBs (e.g., 39 kpc in projection for GRB 230307A, \citealt{Levan:2024}) may be consistent with the merger of a \ac{DNS} in which the second neutron star originated from a particularly massive pre-supernova progenitor, providing a larger systemic kick to the binary.  However, in view of very limited statistics and uncertain models, this is, for now, a speculation.  Future observational constraints include a larger data set of long \acp{GRB} with either kilonova observations or strict upper limits precluding supernovae; the properties of the growing sample of Galactic \acp{DNS} with well-measured masses, eccentricities, and Galactic velocities; and the overlaps and gaps between \ac{DNS} mergers observed via gravitational waves, those observed as \acp{GRB}, and those observed as kilonovae.

\section*{Code availability}

Simulations in this paper made use of COMPAS \citep{COMPAS:2021,COMPAS:2022,COMPAS:2025} v3.22.06. 

\section*{Acknowledgements}

IM and PD acknowledge support from the Australian Research Council (ARC) Centre of Excellence for Gravitational Wave Discovery (OzGrav), through project number CE230100016. OS acknowledges support from the Italian National Institute for Astrophysics (INAF) through `Finanziamento per la ricerca fondamentale' grant number 1.05.23.04.04, and also funding by the European Union-Next Generation EU, PRIN 2022 RFF M4C21.1 (202298J7KT - PEACE).  We thank Paz Beniamini, Giuliano Iorio, Hagai Perets, Thomas Tauris and Reinhold Willcox for comments on the manuscript.

\bibliography{Mandel.bib}{}
\bibliographystyle{aasjournal}

\end{document}